\documentclass[apjl]{emulateapj}
\usepackage{amsmath}

\newcommand{\Msolar}{M$_{\odot}$}

\begin{document}

\title{Interrupted Binary Mass Transfer in Star Clusters}
\shorttitle{Interrupted Binary Mass Transfer in Star Clusters}

\author{Nathan W.\ C.\ Leigh$^{a}$}
\affil{Department of Astrophysics, American Museum of Natural History, Central Park West and 79th Street, New York, NY 10024}
\affil{Department of Physics, University of Alberta, CCIS 4-183, Edmonton, AB T6G 2E1, Canada}
\email{$^a$nleigh@amnh.org}

\author{Aaron M.\ Geller$^{b,c}$}
\affil{Center for Interdisciplinary Exploration and Research in Astrophysics (CIERA) and Department of Physics and Astronomy, Northwestern University, 2145 Sheridan Rd, Evanston, IL 60208, USA}
\affil{Department of Astronomy and Astrophysics, University of Chicago, 5640 S. Ellis Avenue, Chicago, IL 60637, USA}
\email{$^b$a-geller@northwestern.edu}
\thanks{$^c$NSF Astronomy and Astrophysics Postdoctoral Fellow}

\author{Silvia Toonen$^{d}$}
\affil{Leiden Observatory, Leiden University, PO Box 9513, NL-2300 RA Leiden, The Netherlands}
\email{$^d$toonen@strw.leidenuniv.nl}

\shortauthors{Leigh, Geller \& Toonen}

\begin{abstract}
Binary mass transfer is at the forefront of some of the most exciting puzzles of modern astrophysics, including Type Ia supernovae, gamma-ray bursts, and the formation of most observed exotic stellar populations.  Typically, the evolution is assumed to proceed in isolation, even in dense stellar environments such as star clusters.  In this paper, we test the validity of this assumption via the analysis of a large grid of binary evolution models simulated with the \texttt{SeBa} code.  For every binary, we calculate analytically the mean time until another single or binary star comes within the mean separation of the mass-transferring binary, and compare this time-scale to the mean time for stable mass transfer to occur.  We then derive the probability for each respective binary to experience a direct dynamical interruption.  The resulting probability distribution can be integrated to give an estimate for the fraction of binaries undergoing mass transfer that are expected to be disrupted as a function of the host cluster properties.  We find that for lower-mass clusters ($\lesssim 10^4$~\Msolar), on the order of a few to a few tens of percent of binaries undergoing mass-transfer are expected to be interrupted by an interloping single, or more often binary, star, over the course of the cluster lifetime, whereas in more massive globular clusters we expect $\ll$ 1\% to be interrupted.  Furthermore, using numerical scattering experiments performed with the \texttt{FEWBODY} code, we show that the probability of interruption increases if perturbative fly-bys are considered as well, by a factor $\sim 2$.
\end{abstract}


\section{Introduction}\label{intro}

Binary mass transfer (MT) is thought to be responsible for the production of 
most observed exotic stellar populations, including blue straggler stars, low-mass X-ray binaries, millisecond pulsars, cataclysmic variables, etc., and the process is thought to contribute significantly to the rate of Type Ia supernovae, gravitational waves, gamma-ray bursts, etc.  A complete understanding of binary evolution is a prerequisite for predicting the rates of these high-energy phenomena.  Many are thought to occur in star clusters, facilitated by their high densities that stimulate direct\footnote{We use the term, \textit{direct}, to indicate that the distance of closest approach between an interloping object and a binary undergoing MT is equal to or less than the binary semi-major axis.} gravitational encounters leading to e.g, exchange interactions, fly-bys and/or collisions.  Such strong encounters can modify the binary orbital parameters, and hence 
directly affect the course of binary evolution \citep[e.g.][]{ivanova:05,hurley:07,marks:11,geller:13a,geller:13b,geller:15,leigh:12,leigh:13,leigh:15}.     

By necessity, full star cluster dynamical simulations, including both direct $N$-body \citep[e.g., \texttt{NBODY6}]{aarseth:03} and 
Monte Carlo methods \citep[e.g.][]{spurzem:96,joshi:00,vasiliev:15}, rely on a number of simplifying assumptions in order to treat binary star evolution.  In particular, binary MT is often parameterized, following a pre-calculated grid of more detailed models that were evolved in isolation.  These models are based on a sophisticated treatment of stellar evolution, with some basic assumptions for the actual MT process (e.g. energy conservation, mass conservation, etc.).  The hydrodynamics involved in the MT process is traditionally not modeled directly within full star cluster simulations.  Moreover, even if the parameterizations are able to approximate the binary evolution accurately, perturbations during MT are rarely treated properly (and are often ignored).  For example, a direct encounter involving a binary undergoing MT could disrupt the binary or exchange a different star into the system, thus halting the MT process.  Scenarios such as these are not accounted for in many star cluster dynamical models.

The assumption that MT occurs in relative isolation (even in star clusters) is often justified by comparing time-scales; in many clusters the duration of MT, particularly for high-mass stars, can be quite short when compared to the single - binary (1+2) encounter time.  However, for lower-mass stars, the typical duration of MT can greatly increase (e.g., due to the longer thermal and nuclear time-scales within the stars).  Furthermore, the binary - binary (2+2) encounter time-scale can be much shorter than the 1+2 encounter time in clusters with large binary frequencies.\footnote{The exact binary fraction at which the 2+2 encounter rate begins to exceed the 1+2 rate is around $f_\text{b} \sim 10\%$ \citep[e.g.][]{sigurdsson93,leigh:11}.}

In this paper, we begin to quantify the validity of the assumption that MT can be treated as an isolated process within star clusters.  Due to mass segregation and dynamical friction, 
most binaries live in the dense cluster core, where they may not be allowed to evolve fully before being interrupted.  We use the binary evolution code \texttt{SeBa} \citep{portegieszwart96,nelemans01,toonen12} 
to simulate the time evolution of the binary orbital parameters of individual binaries undergoing stable MT, as described in Section~\ref{method}.  We then compare the MT time-scales to analytic estimates for the time-scales over which other single and binary stars in the cluster are expected to encounter, or ``interrupt'', the ongoing MT, as calculated in Section~\ref{times}.  We show that the fraction of interrupted binaries can reach a few tens of percent, for certain cluster parameters.  Finally, we discuss in Sections~\ref{discussion} and~\ref{conc} the implications of our results and offer some concluding remarks.

\section{Binary Evolution} \label{method} 
 
We present results from $\sim10^6$ individual binary evolution simulations performed using the 
\texttt{SeBa} code.  \texttt{SeBa} is a fast stellar and binary evolution code, that parameterizes stars by their mass, radius, luminosity, core mass, etc. as functions of time and initial mass.  The binary evolution includes mass loss, MT, angular momentum loss, common envelope evolution, magnetic braking and gravitational radiation. We consider only stable MT between hydrogen-rich non-degenerate stars. The stability of MT and the MT rate depend mainly on the adiabatic and thermal response of the donor star's radius, and the response of the Roche lobe, to the re-arrangement of mass and angular momentum within the binary. For an overview of the method and a comparison with other methods, see \citet{toonen14}, Appendix B.  

We follow a nearly identical method to define the masses and orbital parameters of our initial binary population as in \citet{geller:15}.  Briefly, primary masses are chosen from a \citet{kroupa:93} initial mass function between 0.1~\Msolar and 100~\Msolar. We draw mass ratios ($q=m_2/m_1$) from a uniform distribution such that $q\leq1$ and $m_2>0.1$~\Msolar.  We choose binary orbital periods and eccentricities from the log-normal and flat distributions, respectively, as observed by \citet{raghavan:10} for binaries with solar-type primary stars in the Galactic field, which are also consistent with observations of solar-type binaries in open clusters (OCs) \citep[e.g.,][]{geller:10,geller:13a,geller:12}.   

We initially performed a set of simulations allowing binaries to occupy nearly the entire log-normal period distribution
and determined that the longest-period binary that will undergo MT in \texttt{SeBa}
within a Hubble time (over all primary masses) is $\log(P\text{[days]}) = 4.16$.  We then imposed this maximum orbital period, along with a minimum orbital period chosen such that the binaries are initially detached, and drew $10^6$ random binaries for \texttt{SeBa}.\footnote{This number of binaries is chosen to be sufficiently large to provide reliable statistics.}  Binaries in the contact stage are excluded in our analysis.

We performed two sets of experiments, one at $\text{[Fe/H]}=-1.5$, typical for an old globular cluster (GC), and the other at $\text{[Fe/H]}=0$, appropriate for an OC.  We use the \texttt{SeBa} code to evolve these binaries for a Hubble time and produce time-averaged orbital parameters (i.e., mean separation, mean binary mass) calculated over the course of stable MT, as well as the cluster age at the onset of MT and the total duration of the MT phase, $t_\text{d}$, for each binary.  

\section{Encounter Time-scales} \label{times}

We use the output from \texttt{SeBa} to calculate the expected time-scales for another single or binary star in the cluster to (directly) encounter, or ``interrupt'', each of the binaries while MT is ongoing, which we will call $t_\text{e}$.  We then compare this time-scale directly to the MT duration, $t_\text{d}$, which is calculated and outputted by \texttt{SeBa}.

To calculate the encounter time-scale, $t_\text{e}$, we must first find the time for a single ($_1$) or binary ($_2$) star to interrupt an ongoing episode of stable MT, which we will refer to as, respectively, $\tau_{2+1}$ and  $\tau_{2+2}$.  As in, \citet{geller15b}, we define these timescales as follows \citep{leonard:89,leigh:11}:
\begin{align} \label{eq:tau_enc1}
\beta &= \left(\frac{10^3\text{ pc}^{-3}}{n_0}\right) \left(\frac{v_\text{rms,0}}{5\text{ km s}^{-1}}\right) \left(\frac{0.5\text{ M}_{\odot}}{m}\right) \left(\frac{1\text{ AU}}{R_\text{enc}}\right)\text{ ,} \nonumber \\
\tau_{2+1} &= 1.4 \times 10^{11} \left(1 - f_\text{b}\right)^{-1} \beta\text{ yr ,} \nonumber \\
\tau_{2+2} &= 5.4 \times 10^{10} \left(f_\text{b}\right)^{-1} \beta\text{ yr ,} \nonumber \\
\end{align}
where $n_0$ is the star cluster central number density, $v_\text{rms,0}$ is the central root-mean-square velocity 
(and we assume $v_\text{rms,0}=\sigma_0=\sqrt{3}\sigma_\text{0,1D}$),
$m$ is the mean mass of a star in the cluster (for which we adopt $m$ $=$ 0.5~\Msolar), $f_\text{b}$ is the binary frequency, and $\pi$$R_\text{enc}^2$ is the mean gravitationally-focused cross section, where $R_\text{enc}$ is either the time-averaged separation of the binary undergoing stable MT (for 1+2 encounters) or the orbital separation corresponding to the cluster hard-soft boundary\footnote{\footnotesize The hard-soft boundary is a theoretical separation in either semi-major axis or orbital period.  Binaries inside this boundary are dynamically ``hard'', and encounters with other stars in the cluster tend to shrink (i.e., harden) the binary.  Binaries beyond this boundary are dynamically ``soft'', and encounters tend to widen and eventually ionize the binary.  Note that this results in an approximate lower limit for the 2+2 time-scale.  We also calculate a rough upper limit below by adopting instead the mean orbital separation below the hard-soft boundary for $R_\text{enc}$.} (for 2+2 encounters).  Finally, the total time until a subsequent encounter is given by,
\begin{equation}  \label{eq:tau_enc}
t_\text{e} = \left(\Gamma_{2+1} + \Gamma_{2+2}\right)^{-1} \\
\end{equation}
where $\Gamma=1/\tau$.

To calculate the parameters in Equation~\ref{eq:tau_enc1}, we consider total cluster masses $M_\text{cl}$ ranging from $10^3$ to $10^6$~\Msolar\ in steps of 1 in $\log_{10}(M_\text{cl}$~[\Msolar]~$)$, 
and assume a Plummer model \citep{plummer:11} with a half-mass radius\footnote{\footnotesize Note that for these parameters, the mean orbital separation during MT is much smaller than the hard-soft boundary, independent of $M_\text{cl}$.} of 3 pc.  See \citet{geller:15} for specific details.  In general $n_0$ increases with increasing cluster mass, while $f_\text{b}$ decreases with increasing cluster mass (as is consistent with observations, e.g., \citealt{leigh:15b}).

\section{Fraction of Interrupted Binaries Undergoing Stable Mass Transfer} \label{frac}

\begin{figure}[!t]
\plotone{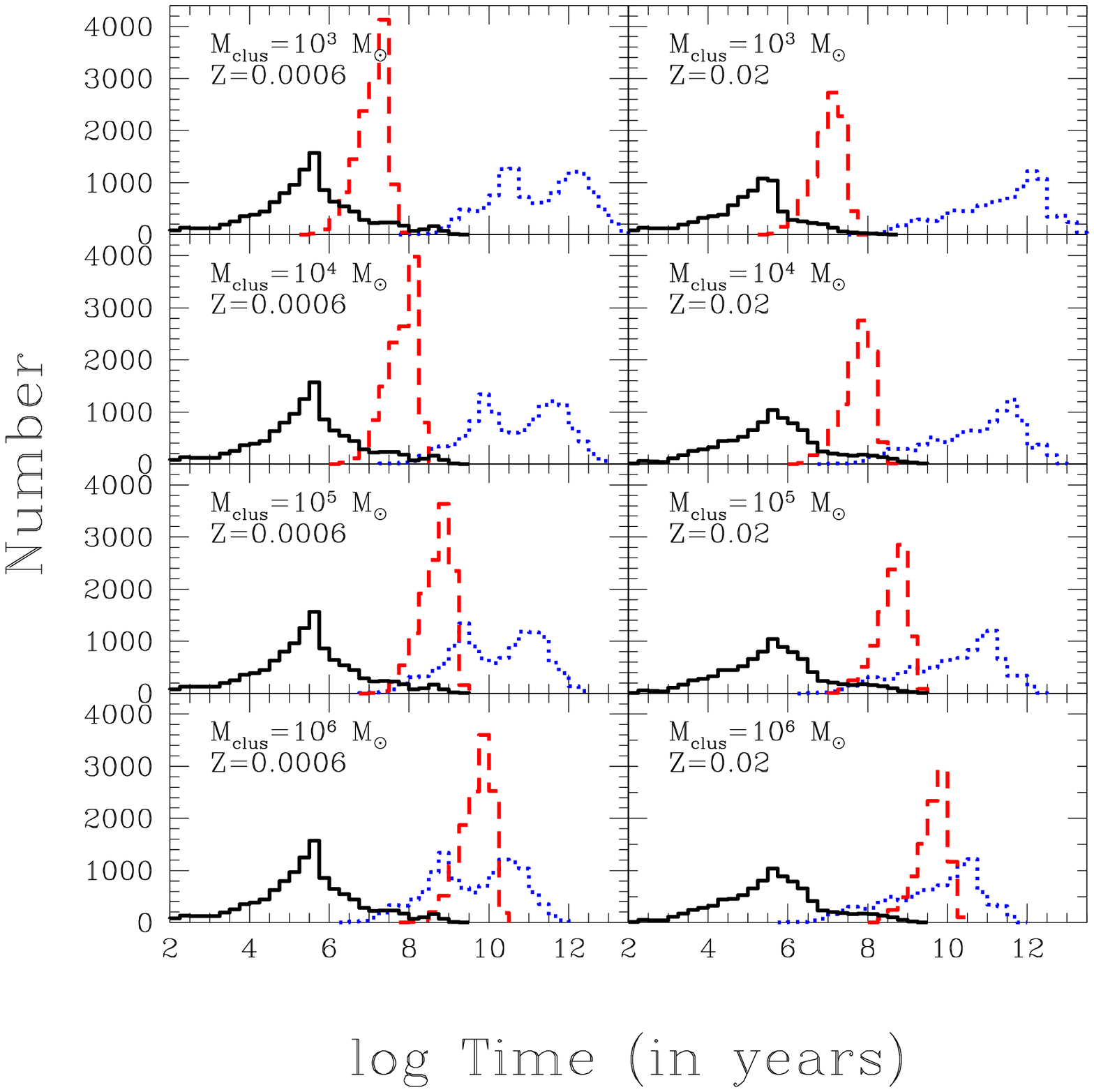}
\caption{
Distributions of the total duration of stable MT ($t_\text{d}$; solid black histograms) and times until the next direct 1+2 ($\tau_\text{1+2}$; dotted blue histograms) and 2+2 ($\tau_\text{2+2}$; dashed red histograms) encounter, drawn from clusters with half-mass radii of 3 pc and the $M_\text{cl}$ values indicated in each panel.  The left and right panels distinguish between assumed metallicities of Z $=$ 0.0006 and Z $=$ 0.02, respectively.
\label{fig:fig1}
}
\end{figure}

\begin{figure}[!t]
\plotone{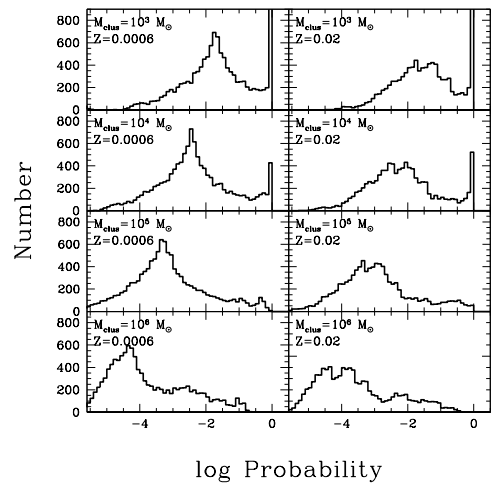}
\caption{
Distributions of the probability for binary MT interruption, as defined in Equation~\ref{eqn:poisson}.  The binary populations used to calculate these probabilities are the same as in Figure~\ref{fig:fig1}.
\label{fig:fig2}
}
\end{figure}

We find from these binary evolution simulations that, the probability that a given binary will be interrupted during MT in a star cluster core ranges from $\ll$ 1\% to well over unity.  The main results are plotted in Figures~\ref{fig:fig1}~and~\ref{fig:fig2}.  

As shown in Figure~\ref{fig:fig1}, both $\tau_\text{1+2}$ and $\tau_\text{2+2}$ are sensitive to the total cluster mass, whereas the MT durations $t_\text{d}$ are not.  The sensitivity of $\tau_\text{1+2}$ and $\tau_\text{2+2}$ to $M_\text{cl}$ is due to the dependence of the semi-major axis at the hard-soft boundary on $M_\text{cl}$, which results (on average) in tighter binaries and lower binary fractions in more massive clusters.  In low-mass clusters, the mean binary separation and binary fraction are both large, causing $\tau_\text{2+2}$ to be short.  As $M_\text{cl}$ increases, both the mean binary separation and binary fraction decrease along with the separation at the hard-soft boundary, causing $\tau_\text{1+2}$ and $\tau_\text{2+2}$ to decrease and increase, respectively.  Importantly, most binaries undergoing stable MT have time-averaged separations that are on the order of only a few stellar radii, causing $\tau_\text{1+2}$ to exceed a Hubble Time nearly independent of $M_\text{cl}$.  On the other hand, $\tau_\text{2+2}$ is always much shorter than a Hubble Time, due to the larger mean separation of the entire cluster binary population.

In Figure~\ref{fig:fig1}, a bimodality is apparent in the distribution for $\tau_\text{1+2}$ at low metallicity (i.e. Z $=$ 0.0006) that is not apparent in either the $\tau_\text{1+2}$ distribution at high metallicity (i.e. Z $=$ 0.02) or in either of the $\tau_\text{2+2}$ distributions.  This is due to a bimodality that appears in the mean binary separation during mass transfer, which dominates the 1+2 cross-section but not the 2+2 cross-section, since the mean binary separation in the cluster is much larger than the mean binary separation during mass transfer.  The first peak in this bimodal distribution (at small orbital separations) is due to systems for which stable mass transfer is initiated in a relatively tight binary, involving relatively unevolved stars. The second peak is due to systems with asymptotic giant branch (AGB) donors that have lost so much mass that they are less massive than their companion.  Consequently, the resulting mass transfer is stable.  For low metallicities, this second group is bigger, and a second peak becomes clearly visible (relative to high metallicities).  This is because the lifetimes of low metallicity stars are shorter relative to more metal-rich stars.  Therefore, low mass stars are able to evolve off the main sequence, become an AGB star, and hence fill their Roche lobe all within a Hubble time.

In Figure~\ref{fig:fig2}, we compare $t_\text{d}$ and $t_\text{e}$ for each individual encounter directly, by calculating the cumulative Poisson probability or the probability that a given binary will undergo at least one encounter with another star or binary within a time interval $t_\text{d}$:
\begin{equation}
\label{eqn:poisson}
P = 1 - e^{-t_\text{d}/t_\text{e}}
\end{equation} 
Equation~\ref{eqn:poisson} approximates the probability that a given binary undergoes a \textit{direct} encounter with another star or binary in the cluster before MT stops (shown on the x-axis in Figure~\ref{fig:fig2}).

Importantly, fly-by encounters will also perturb binaries undergoing MT, and this is not accounted for in Figures~\ref{fig:fig1} and~\ref{fig:fig2}.  To quantify this effect, we ran simulations of 1+2 encounters using the \texttt{FEWBODY} code \citep{fregeau:04}.  We fix the binary orbital parameters and relative velocity at infinity for the encounters, but sample the impact parameter in multiples of 0.5$b_\text{0}$, where $b_\text{0}$ is the impact parameter corresponding to a distance of closest approach equal to the binary semi-major axis $a$, or \citep{leonard:89,leigh:11}:
\begin{equation}
\label{eqn:imp0}
b_\text{0} = \Big( \frac{6Gma}{v_\text{inf}^2} \Big)^{1/2}
\end{equation}
Here, $m$ is the mean single star mass in the cluster and $v_\text{inf}$ is the relative velocity at infinity.  We perform $10^4$ numerical scattering experiments for every choice of the impact parameter.  All stars are assumed to be point particles with masses of 1\Msolar.  The initial binary semi-major axis and eccentricity are 1 AU and 0, respectively.  The incoming velocity at infinity is set to $v_\text{inf} = 5$ km/s.  

\begin{figure}[!t]
\plotone{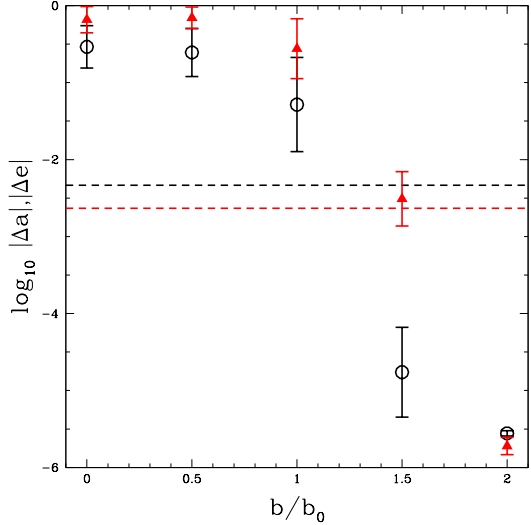}
\caption{The (logarithm of the) mean change in eccentricity |${\Delta}e$| (red triangles) and semi-major axis |$\Delta{a}$| (black open circles; in AU) for binaries involved in single-binary interactions, as a function of the impact parameter (x-axis).  The impact parameter is sampled in units of 0.5$b_{\text{0}}$, defined as in Equation~\ref{eqn:imp0}.  The uncertainties correspond to the standard deviation of the mean. (The distributions for |$\Delta{a}$| and |$\Delta{e}$| are not strictly Gaussian, so the uncertainties should formally be asymmetric.  However the differences are negligible for our purposes here.)   The dashed lines correspond to the critical values of $|\Delta{a}|$ (black) and $|\Delta{e}|$ (red; in AU) defined by our criteria for a "significant" perturbation described in the text. 
\label{fig:fig3}
}
\end{figure}

The results of these scattering experiments are illustrated in Figure~\ref{fig:fig3}, which shows the mean (absolute) change in the binary semi-major axis and eccentricity for all simulations, as a function of the impact parameter.  We assume here that a fly-by encounter ``interrupts'' a given binary undergoing MT if either of the two following criteria are met: 1) the change in semi-major axis $|\Delta{a}|$ is greater than or equal to the typical stellar radius at the onset of Roche lobe overflow ($\sim$ 1 R$_{\odot}$ for most binaries), and/or 2) the change in eccentricity $|\Delta{e}|$ corresponds to a change in the quantity $r_\text{p} - r_\text{a}$ that is greater than or equal to the typical stellar radius at the onset of MT, where $r_\text{p}$ and $r_\text{a}$ are the pericenter and apocenter distances, respectively, from the binary center of mass.  We assume that the magnitude of such a perturbation is sufficiently large to at least temporarily halt the MT process (provided $\Delta{a} > 0$ and/or $\Delta{e} < 0$).    

As shown in Figure~\ref{fig:fig3}, the perturbative effects of fly-bys remain significant for b $\lesssim$ 1.5$b_\text{0}$.
Moreover, both $\left|\Delta{a}\right|$ and $\left|\Delta{e}\right|$ undergo a steep drop-off between b $=$ 1$b_\text{0}$ and b $=$ 2$b_\text{0}$, such that neither of our criteria for a ``significant'' perturbation are satisfied 
for b $\gtrsim$ 1.5$b_\text{0}$.  Our analytic estimates for the 1+2 and 2+2 encounter rates are directly proportional to the gravitationally-focused cross-section $\sigma_\text{gf} \propto {\pi}b^2$, for some 
impact parameter b.  Hence, if fly-bys are included, the time-scales for 1+2 and 2+2 encounters should each decrease by a factor $\lesssim 1.5^2 \sim 2.1$, as should $t_\text{e}$.  Working from Equation~\ref{eqn:poisson}, we find that the probabilities typically increase by a corresponding factor of $\sim$ 2.  That is, the ratio between the probabilities calculated with and without this additional factor of $\sim$ 2.1 in our estimate for $t_\text{e}$ peak at a value of $\sim$ 2 with very little scatter. 

\begin{figure}[!t]
\plotone{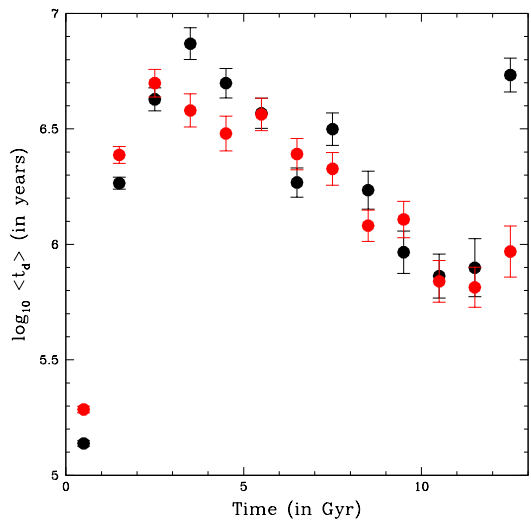}
\caption{The (logarithm of the) mean duration of mass transfer (shown on the y-axis; in years) as a function of the (logarithm of the) host cluster age (x-axis; in Gyr), for all cluster masses adopted in Figures~\ref{fig:fig1} and~\ref{fig:fig2}.  The mean is calculated in 1 Gyr intervals, using all binaries that initiate mass transfer during each 1 Gyr window of the cluster lifetime.  Error bars correspond to the standard error of the mean.  Black and red solid circles correspond to metallicities of, respectively, Z $=$ 0.0006 and Z $=$ 0.02. 
\label{fig:fig4}
}
\end{figure}

The fraction of binaries expected to be interrupted during MT can be estimated from the probability distributions presented in Figure~\ref{fig:fig2}.  To do so, we integrate the probability distributions in Figure~\ref{fig:fig2} and divide by the total number of binaries.  The resulting fractions are $f_\text{i} \sim$ 0.18, 0.10, 0.03 and 0.01 for our model clusters with $M_\text{cl}$[\Msolar] $=$  10$^3$, 10$^4$, 10$^5$ and 10$^6$, respectively, and a metallicity Z $=$ 0.02.  These fractions remain the same to within $\pm 1$\% for a metallicity Z $=$ 0.0006.  As shown in Figure~\ref{fig:fig4}, the mean mass transfer duration is $\sim$ 10$^{5.5}$ years during the first 1 Gyr of cluster evolution, then quickly increases by roughly an order of magnitude.  After this, the mean MT duration slightly and slowly decreases over the next $\sim$ 12 Gyr, but remains much larger than during the initial 1 Gyr.  Hence, if the initial 1 Gyr of the cluster lifetime is excluded from our calculation for the fraction of binaries expected to be interrupted during MT, these estimates increase by a factor $\sim$ 2.\footnote{Note that the final increase in the mean duration of mass transfer at $\sim$ 13 Gyr seen for the Z $=$ 0.0006 case should be regarded with caution.  This is due in part to small number statistics, and the presence of a handful of binaries with unusually long MT durations at late times.  This is not visible in the error bars, since we are showing the standard error of the mean here, and the number of binaries included in this calculation is large.  Perhaps more importantly, these error bars do not reflect any inherent uncertainties in the \texttt{SeBa} code and binary evolution in general, which are certainly non-negligible.}  These fractions correspond approximately to upper limits, since we adopt the mean orbital separation at the hard-soft boundary when calculating the 2+2 encounter time.  If we adopt instead the mean orbital separation \textit{below} the hard-soft boundary (which does not include the presence of soft binaries), these fractions decrease by a factor of a few (of the same order as the increase from including fly-bys).

From these fractions, we can also estimate the number of binaries that are expected to be directly interrupted during ongoing stable MT:
\begin{equation}
\label{nbin}
N_\text{bin,i} = N_\text{bin}f_\text{MT}f_\text{i}, 
\end{equation}
where $N_\text{bin}$ is the total number of initial binaries in the cluster (estimated as $f_\text{b}N_\text{stars}$), $f_\text{MT}$ is the fraction of binaries expected to undergo MT (calculated by \texttt{SeBa} for our assumed binary orbital parameter distributions) and $f_\text{i}$ is the fraction of binaries undergoing MT that are expected to be interrupted over a Hubble time of cluster evolution (as calculated above for our four cluster masses).  These parameters are shown in Table~\ref{table:one}, for all four assumed values of the total cluster mass.

\begin{table}
\begin{center}
\caption{Parameters adopted in Equation 5 to calculate the numbers of interrupted MT binaries.}
\begin{tabular}{|c|c|c|c|c|}
\hline
$M_\text{cl}$    &     $N_\text{bin}$   &   $f_\text{MT}$   &    $f_\text{i}$    &    $N_\text{bin,i}$           \\
(in \Msolar)       &                                  &                             &                           &                                           \\
\hline
10$^3$              &        665                   &          0.05           &        0.18           &           6                             \\
10$^4$              &        3837                 &          0.07           &        0.10           &          27                             \\
10$^5$              &       16431                &          0.12           &        0.03           &          59                             \\
10$^6$              &       50338                &          0.23           &        0.01           &         116                              \\
\hline
\end{tabular}
\label{table:one}
\end{center}
\end{table}

The resulting total numbers of binaries expected to be interrupted over the lifetime of the clusters is then $N_\text{bin,i} \sim$ 6, 27, 59 and 116 for our model clusters with $M_\text{cl}$[\Msolar] $=$  10$^3$, 10$^4$, 10$^5$ and 10$^6$, respectively.  If perturbative fly-bys are included, these values each increase by a factor $\sim$ 2.  \textit{The calculated rate of MT interruption for clusters is $\sim$ 1-10 binary per Gyr, with only a weak tendency for this rate to increase with increasing cluster mass.}

\section{Discussion} \label{discussion}

We find that a relatively large fraction $f_\text{i}$ of binaries undergoing stable MT may encounter another single or binary star in the core of a star cluster before the MT proceeds to completion.  We calculate $f_\text{i} \sim$ 0.18, 0.10, 0.03 and 0.01 for our model clusters with $M_\text{cl}$[\Msolar] $=$  10$^3$, 10$^4$, 10$^5$ and 10$^6$, respectively, nearly independent of metallicity.  These fractions are only slightly lower than we calculated in \citet{geller15b} for the fraction of interrupted stellar encounters in star clusters, which ranges from $<$ 0.01 in massive GCs to $>$ 0.4 in low-mass OCs.  The origin of this analogous dependence between the fraction of interrupted "events" and the total cluster mass is ultimately the same (see below).  

As a binary undergoing MT generally has a relatively small geometric cross section, the interloper in these cases are most often binary stars themselves, resulting in 4-body encounters that may dramatically alter the course of the binary evolution.  Interrupted MT is expected to occur most often in massive OCs and low-mass GCs ($\sim 10^4$~\Msolar), where in general the binary frequency is larger, and (detached) binaries can have larger orbital separations and therefore larger geometric cross sections.  These larger binary frequencies and geometric cross sections present in the low-mass cluster regime also result in a larger fraction of interrupted stellar encounters.  No publicly available code for star cluster dynamics accounts directly for the hydrodynamics of binary MT needed to accurately model such interrupted/perturbed MT binaries.

As discussed in \citet{geller15b} for interrupted stellar encounters, the dependence of $f_\text{i}$ on $M_\text{c}$ is encouraging for numerical models of star clusters, since low-mass clusters such as OCs are most often modeled
with direct $N$-body simulations.  Direct $N$-body codes, such as \texttt{NBODY6}, can in principle apply perturbations to binaries undergoing (parameterized) MT, if the time steps are chosen appropriately.  Given our results, a more detailed study of how often such cases of interrupted MT occur and how accurately they are treated in direct $N$-body codes is warranted.  Massive clusters, on the other hand, are more often modeled with Monte Carlo codes \citep[e.g.,][]{chatterjee:10,hypki:13}.  Here, binary evolution is assumed to run to completion in isolation.  In massive GCs, we find that this assumption is much more valid than in OCs (albeit not perfect, see Figure~\ref{fig:fig2}), similar to what was found in \citet{geller15b} for interrupted stellar encounters.  

The time-averaged separations of binaries undergoing stable MT are typically on the order of a few solar radii.  However, stable and conservative MT leads to an increase in the binary orbital separation (if the donor is less massive than the accretor).  Generally toward the end of the MT phase,
after a mass-ratio inversion, the orbits expand significantly, increasing the binary cross-section and decreasing the 1+2 and 2+2 encounter times.  Hence, the time-scale for interruption decreases markedly at the end of the MT phase and beyond, which is not properly accounted for in our models (since we only use the mean binary separation during MT).  Interestingly, the final separation is on the order of $\gtrsim$ 1 AU in our models, which is comparable to the hard-soft boundary in old massive ($\sim 10^6$~\Msolar) GCs.  Hence, these binaries become subject to ionization due to dynamical encounters with other single and binary stars after MT.  Assuming a MT origin, this could be of particular relevance for blue straggler formation in GCs.  In particular, this naively predicts a rough correlation whereby the fraction of blue stragglers in binaries increases with decreasing $M_\text{cl}$.

\section{Conclusions} \label{conc}

Even the very intimate dance of binary MT can be interrupted by an interloping single or, more often, binary star in a star cluster.  Up to about a quarter of all binaries undergoing stable MT in the cores of massive OCs and low-mass GCs ($\sim 10^4$~\Msolar) may encounter another star (or stars) before completing this phase of evolution (see Figure~\ref{fig:fig2}).  This fraction decreases towards $<<1$\% for more massive GCs, primarily because the binary frequency in these clusters is much lower.  Cases of interrupted MT may lead to significantly different outcomes than are assumed for isolated evolution, which in turn may impact the formation rates of MT products in star clusters, from exotic stars like blue stragglers, to SN Ia and gamma-ray burst progenitors. Our results highlight the need for a more detailed treatment of binary evolution in star cluster simulations.

\acknowledgments
The authors would like to thank an anonymous reviewer for useful suggestions for improvement.  N.W.C.L.\ is grateful for the generous support of an NSERC Postdoctoral Fellowship.  A.M.G.\ is funded by a National Science Foundation Astronomy and Astrophysics Postdoctoral Fellowship under Award No.\ AST-1302765.

\bibliographystyle{apj}

\end{document}